# Pyroelectric charge release in rhombohedral PZT.

Beatriz Noheda, Ning Duan*, Noé Cereceda and Julio A. Gonzalo.
*Dept. Física de Materiales, C-IV. Universidad Autónoma de Madrid. Cantoblanco. 28049-Madrid. Spain.*
* On leave from Shanghai Inst. of Ceramics.

A new experimental set-up controlled by computer has been made to measure the pyroelectric charge of ferroelectric materials with a relatively high conductivity at slow rates of temperature variation. It allowed us to obtain the polarisation vs. temperature behaviour of PZT with various compositions in the whole range of temperatures with a high accuracy. The ferroelectric Zr-rich region of the PZT phase diagram, where the isostructural phase transition $F_{RL}$-$F_{RH}$ takes place, has been studied.

## I. INTRODUCTION

The isostructural phase transition between two ferroelectric rhombohedral phases ($F_{RL}$ and $F_{RH}$) of PZT is gathering increasing interest due to its large pyroelectric figure of merit, caused by its high pyroelectric coefficient combined with a low dielectric constant[1]. From the structural point of view, the tilt of the oxygen octahedra in the low temperature phase ($F_{RL}$) is responsible for the phase transition[2]. The tilt is coupled with the polarization in Zr-rich compositions, giving place to small anomalies in the polarization and the dielectric constant behaviour[3]. The disappearance of these discontinuities with increasing Ti content has been recently discussed by Viehland et al.[4] in thin plates, proposing different kinds of short-range oxygen rotations also in the ferroelectric high temperature phase ($F_{RH}$). Nevertheless there are other factors, such as the compositional disorder, which makes the transition broader and the anomalies smaller, and could contribute to the vanishing of the anomalies at the $F_{RL}$-$F_{RH}$ transition as the Ti content increases.

The tilt-polarization coupling has been studied by the authors by means of neutron diffraction and hysteresis loops data[3]. Nevertheless, the available experimental data on spontaneous polarisation ($P_s$) vs. temperature were scarce and not accurate, due to the experimental difficulties. To obtain better experimental data has been the main motivation of this work.

It is not easy to characterise dielectrically, as a function of temperature, ferroelectric ceramics such as PZT and similar. The low impedance found in these materials close to the ferroelectric transition makes very difficult the right compensation of the P-E hysteresis loops. The difficulties increases at low frequencies. These materials are insulators at room temperature but they become notably semiconductors at temperatures close to the ferroelectric phase transition (200-300 ºC). The high increment of the conductivity produces strong variations with temperature of the optimum electric compensation in the measurement circuit, which gives rise unreliable results. The same problem is found measuring the dielectric permitivity, where the high temperature response is very often masked by the conductivity. On the other hand, the high coercive fields existing far bellow from the ferroelectric transition hinder the observation of the hysteresis loops.

Polarization measurements in rhombohedral PZT by P-E loops have both above mentioned problems: high coercive fields and low impedance. Due to the fact that the ferroelectric ($F_{RL}$)-ferroelectric ($F_{RH}$) phase transition occurs close to room temperature, where the high coercive fields avoid the P-E loops, it is one case where a different procedure to determine spontaneous polarization is really necessary.

Pyroelectric charge measurements are an excellent alternative to the hysteresis loops. If accurate temperature determination is required (slow measurements), using an operational amplifier (OA) as charge integrator, is necessary to keep the sample at zero electric field. In this way the resulting charge integration is independent of the sample impedance. This consideration led to the system used by Bernard et al.[5] to study rhombohedral PZT. An improved version of such a system has been developed in this work and it has been used to investigate Zr-rich PZT.

## II. EXPERIMENTAL

The experimental set-up is based, as it has been pointed out, in that used by Bernard et al.[5], which allows polarization and charge integration, alternatively, by means of two switches. The system has three main parts: a high voltage variable source, a charge integrator and a temperature sensor . The switches are computer controlled and the data are automatically recorded.

The use of charge integrators with ferroelectric materials is frequently avoided because of the extreme fragility of the operational amplifier (OA) needed. These devices must be capable of integrating very low currents, of the order of nA's, which implies bias currents of the order of fA's. But, at the same time, they are very often exposed to short-circuits resulting in high voltages, which can destroy the OA. A substantial improvement in the system has consisted in carefully protecting the device against high voltage peaks. This, however, is not a trivial accomplishment in our case because p-n unions, currently used to this end, produce offset currents of nA's, of the same order of magnitude as that of the currents to be measured. This problem has been solved here using an specific FET transistor which limits the maximum voltage to be applied to the AO, giving, at the same time, offset currents in the range of fA. The integrator scheme is shown in Fig. 1.

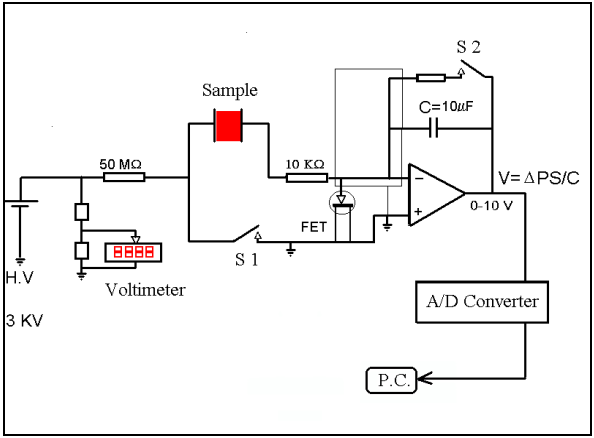

FIG. 1. Scheme of the charge integrator developed to measure pyroelectric charge in ferroelectric materials with relatively high conductivity at slow rates of temperature variation.

The working procedure begins with the poling of the sample somewhat above the $F_{RL}$-$F_{RH}$ phase transition, opening the "poling switch", S1, (see Fig.1). Sample poling was performed with a field of 20 kV/cm during 15 min. in a silicon-oil bath at 125ºC. After poling, the sample is taken to the lowest temperature through the $F_{RL}$-$F_{RH}$ phase transition and then the temperature is increased again to 150ºC, without integrating the pyroelectric charge during this first temperature cycle. At this point the charge integration starts, opening the "integration switch", S2, which has been closed until this moment to allow the discharge of the measurement capacitor, C. During integration, the sample is first cooled through $F_{RL}$-$F_{RH}$ transition (at $T_{LH}$) and, second, heated to the paraelectric phase. Operating in such a way the thermal hysteresis of the $F_{RL}$-$F_{RH}$ transition can be observed. The measured voltage is the OA output, which is proportional to the charge released, $V= \Delta Q/C= \Delta P_s S/C$, being $P_s$ the spontaneous polarization and S the samples electroded surface. The constant voltage level at the paraelectric phase serves to determine the reference point to get the absolute value of the spontaneous polarization. The rate of temperature change used was 20 ºC/h in each case.

Samples were ceramic discs of 1mm in thickness and 15 mm in diameter for nine different compositions with Zr/Ti ratios, (100-X)/X, between 98/2 and 77/23. Sample preparation is described in ref. 3. The sample conductivity was controlled in part adding 1% wt of $Nb_2O_5$.

### III. RESULTS

$P_s$ vs. temperature curves are shown in Fig.2 for the nine compositions studied. The spontaneous polarization has been normalised dividing by $P_{so}= P_s(T= 0K)$, obtained fitting the data as in ref. 3. Both the ferroelectric ($F_{RH}$-$P_C$) and the isostructural ($F_{RL}$-$F_{RH}$) transitions are shown.

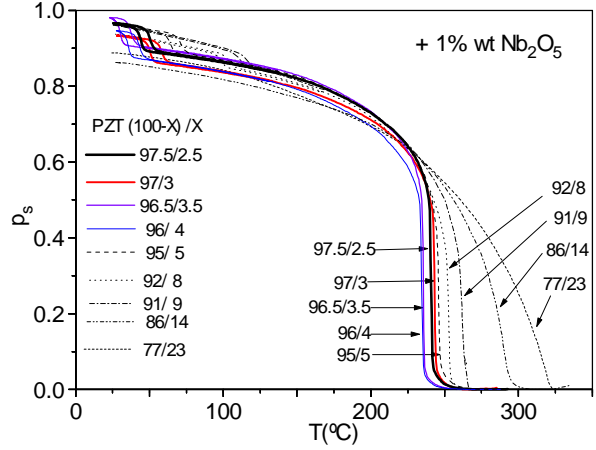

FIG.2. Normalised spontaneous polarization ($p_s= P_s/P_{so}$) versus temperature of PZT(100-X)/X with 1 % by weight of $Nb_2O_5$, for nine different compositions in the rhombohedral region.

The first order character of the ferroelectric ($F_{RH}$-$P_C$) transition at $T_o$ is clearly observed at the higher Zr-content. This character is decreasing with increasing Ti content. It has been found (not shown in Fig 2 for clarity) that the conductivity has a large effect above temperatures going from 270 to 300ºC. $P_s$ vs. T curves are far from being flat above $T_o$ for compositions with X< 10, which, at these temperatures, are in the paraelectric phase. This problem leads to consider possible distortions in the polarization curve above 270ºC for compositions (X>10) still ferroelectric at these temperatures, as the case for PZT 86/14 and 77/23. Due to this fact, the tricritical point, at which the transition character changes from first to second order, is difficult to determine experimentally.

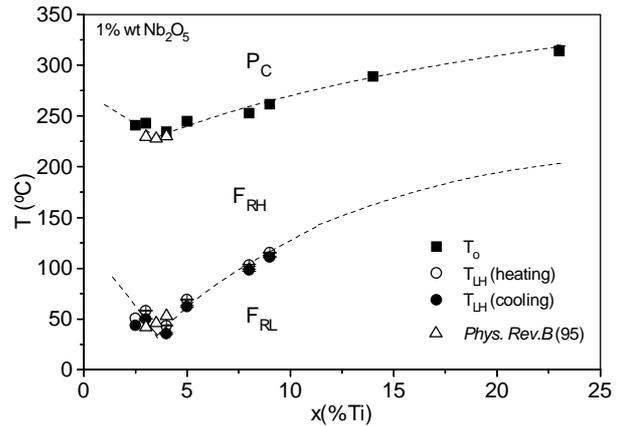

FIG. 3. Phase diagram of PZT(100-X)/X + 1%wt $Nb_2O_5$ (0<X<25) from pyroelectric charge data shown in Fig. 2. Previously reported data are also included.

Heating and cooling through the $F_{RL}$-$F_{RH}$ transition ($T_{LH}$) allows the determination of the thermal hysteresis and the first order character of the isostructural transition for compositions with higher Zr content.

The phase diagram obtained from the data in Fig.2 is plotted in Fig.3. Data from previous work[6] are also included. It can be observed that the $F_{RL}$-$F_{RH}$ phase boundary is shifted to the left, by a constant value of 2.6% along the X axis, with respect to the phase diagram of pure PZT[7]. The value of the shift corresponds, in this range of compositions, to the molar content of Nb added, which shows that the Nb addition does not modify the $F_{RL}$-$F_{RH}$ transition behaviour. This supports the idea of $Nb^{5+}$ being located at $Zr^{4+}$site, and not in interstitial positions[8,9].

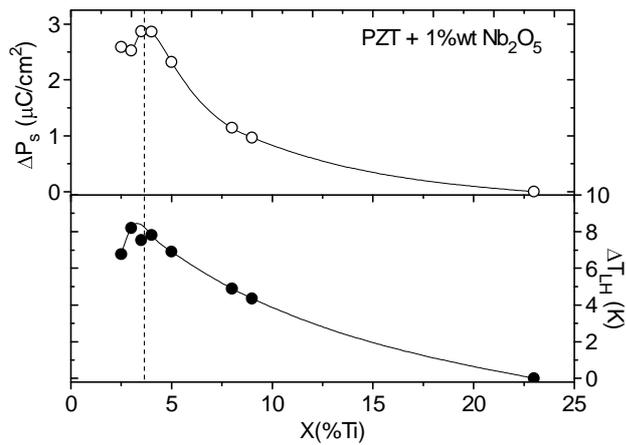

FIG 4. Change in spontaneous polarization ($\Delta P_s$) and thermal hysteresis ($\Delta T_{LH}$) of the $F_{RL}$-$F_{RH}$ transition versus Ti content (X) for PZT with 1% wt of $Nb_2O_5$.

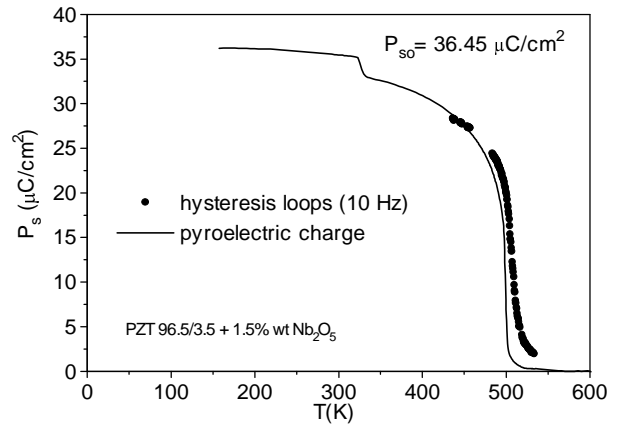

FIG.5. Spontaneous polarization vs. T. by integrating the pyroelectric charge measured also at low temperatures, to determine $P_{so}= P_s(0 \text{ K})$. Hysteresis loops data are shown to illustrate the experimental difficulties of this method.

Thermal hysteresis, $\Delta T_{LH}= T_{LH}(\text{heat.})-T_{LH}(\text{cool.})$, and increment of the spontaneous polarization ($\Delta P_s$) both measured through the $F_{RL}$-$F_{RH}$ transition are shown in Fig. 4. $\Delta T_{LH}$ and $\Delta P_s$, decrease with increasing Ti content. The $P_s$ step occurring at $T_{LH}$ becomes less sharp, it diffuses and broadens over a wider range of temperature as the Ti content is increased, becoming undetectable for the compositions examined with Zr/Ti ratio lower than 86/14. Something happens, however, for Zr/Ti above 96/4. These compositions show clear ferroelectric behaviour above room temperature due to the Nb diminishing the antiferroelectric region. Their Ti content (X< 4% molar) is smaller or similar to the Nb added (2.6% molar). This reversal in trend of $T_{LH}(X)$ is clearly at variance with that shown by the data in Jaffe's phase diagram for pure PZT.

A modification of the set-up to measure in the low temperature range (160 K-370 K) has been developed later and it has allowed us to examine the experimental value of $P_{so}$, as it is shown in Fig.5 for PZT 96.5/3.5+1.5%wt $Nb_2O_5$.

In this figure we also show data obtained in the same sample by hysteresis loops measurements, with an available applied field of 10 kV/cm at a frequency of 10 Hz, to illustrate the experimental difficulties found with this method. Apart from the hysteresis loops dependence on frequency, and the lack of field amplitude to saturate the loop far from the paraelectric phase, the bad compensation of the loops caused by the conductivity, leads to a broadened phase transition, which is, however, well define by pyroelectric measurements.

In conclusion, a specially designed experimental set-up, based on one used by Bernard et al., has been developed to measure pyroelectric response of low impedance ferroelectric materials at temperature rates as slow as one may wish. It has allowed to measure accurately the thermal hysteresis and the discontinuous step of the spontaneous polarization at the $F_{RL}$-$F_{RH}$ transition of Nb-doped Zr-rich PZT ceramics. The phase diagram of PZT with 1%wt of $Nb_2O_5$ for 300 K< T< 600 K and 2% < X< 25% has been carefully established.

## V. ACKNOWLEDGEMENTS

Our thanks to Dr. C. Alemany for his helpful comments and suggestions. The financial support of CICyT (PB96-0037) and NATO is acknowledged (CGR-0037).